%% file: main.tex
\documentclass[twocolumn]{aa}  

\usepackage{amsmath}
\usepackage{graphicx}
\usepackage{tikz}           
\usepackage{changes}
\usepackage{txfonts}
\usepackage{bm}
\usepackage{color,soul}
\usepackage{longtable}
\usepackage{pgffor} 
\usepackage{xspace} 

\usepackage{hyperref} 

\newcommand{\tempus}{\textsc{Tempus}\xspace}

\def\ODeux{\mathcal{O}\left( {c^{-2}} \right)}
\def\OQuatre{\mathcal{O}\left( {c^{-4}} \right)}
\def\OCinq{\mathcal{O}\left( {c^{-5}} \right)}

\immediate\write18{bibtex "aa.aux"}

\begin{document}

   \title{TEMPUS: Relativistic coordinate time scales for any solar-system body from arbitrary ephemerides}
   \titlerunning{TEMPUS}

   \author{Y. Seyffert\inst{1} 
          \and
          A. Fienga\inst{1,2} 
          }
    \authorrunning{Seyffert et al.}

   \institute{  Observatoire de la Côte d’Azur, Université Côte d’Azur, CNRS, Géoazur, IRD, 250 avenue A. Einstein 06560 Valbonne, France 
          \and
          Observatoire de la Côte d’Azur, Université Côte d’Azur, CNRS, Artémis, 60 av. de l'Observatoire 06000 Nice Cedex, France
             }

   \date{Received XX; accepted XX}

 
  \abstract
   {Precise time-scale transformations between solar-system bodies are an essential requirement for deep-space science and solar system navigation. The post-Newtonian relativistic expression relating a body-centered coordinate time TCX to the Barycentric Coordinate Time TCB depends on the gravitational environment and velocity of the target body, and must be evaluated consistently with the chosen planetary ephemerides. So far only INPOP has provided such a solution as part of its ephemerides releases, albeit only for Earth with TCG-TCB.}
   {We present \tempus, a tool that numerically integrates the IAU\,2000 time-transformation for any solar-system body using positions and velocities from an arbitrary ephemerides. We provide a validated, ephemerides-agnostic implementation of TCX$-$TCB and assess its accuracy against published independent solutions.}
   {The rate $d\,\mathrm{TCX}/d\,\mathrm{TCB}$ is integrated with an Adams--Moulton order-12 predictor--corrector. The gravitational potential sum tested here includes the Sun, eight planets, Pluto, up to 343 main-belt asteroids, and 30 trans-Neptunian objects. Outputs can be stored in readable format tables of sampled values or as Chebyshev coefficient files. Comparisons are performed against the TU\,Dresden time solutions \citep{Klioner}, the TCG-TCB solution provided with INPOP19a \citep{2019NSTIM.109.....V} and INPOP21a \citep{2021NSTIM.110.....F}, and the LTE440 lunar solution \citep{Lu2025}. We use INPOP19a, INPOP21a, DE430, DE440, and EPM2021 as input ephemerides.}
   {For the planetary bodies \tempus agrees with \cite{Klioner} at the $10^{-21}$\,s\,s$^{-1}$ level over 200\,yr. Against the INPOP19a released TCG-TCB we also agree to the same level. The main-belt asteroids as well as the trans-Neptunian objects shift the planetary time scales at the $10^{-18}$\,s\,s$^{-1}$ level, and the choice of planetary and lunar ephemerides contributes differences on the $10^{-18}$\,s\,s$^{-1}$ level as well.}
   {\tempus provides a validated, ephemeris-agnostic post-Newtonian time-transformation for any solar-system body.}


   \maketitle
   \nolinenumbers

   \input{sections/intro}
   \input{sections/model}
   \input{sections/coords}
   \input{sections/validation}
   \input{sections/perturbers}
   \input{sections/chebyshev}
   \input{sections/applications}
   \input{sections/conclusions}

\begin{acknowledgements}
      We thank S. A. Klioner for fruitful discussions, which deepened our understanding and improved this work, in particular the comparison with his time ephemerides.
      This work was carried in the context of the ESA study “Fundamental Techniques, Models and Algorithms for a Mars Radio Navigation System”, contract No. 4000149060/25/NL/AT.
\end{acknowledgements}

   \bibliographystyle{aa} 
   \bibliography{global}
   
\end{document}

%% file: sections/intro.tex

\section{Introduction}
\label{sec:intro}

According to general relativity, a clock's tick rate depends on its velocity and the gravitational potential at its location: a clock co-moving with the Earth therefore runs at a slightly different rate than an ideal clock co-moving with the Moon, Mars, or solar-system barycenter (SSB). For the Earth this secular drift to the SSB is about $1.28$\,ms per day, and an additional periodic component (driven by orbital eccentricity and planetary perturbations) that must be modeled to nanosecond accuracy for modern space applications.

With the IAU 2000 resolution \cite{Soffel2003}, the IAU gives the metric that should be used for describing the motions of objects in the solar system weak field approximation. Derived from the \cite{Soffel2003} metric and considering each specific gravitational environment, one can define different time-scales: Barycentric Coordinate Time (TCB) at the SSB, its body-mass-centered analogues Geocentric Coordinate Time (TCG), Selenocentric Coordinate Time (TCL), and generically TCX for any body~X at its center of mass, with TCX$-$TCB typically fixed to zero at the 1977 origin, when the SI second came into effect. Closely related to TCB is Barycentric Dynamical Time (TDB), which is TCB rescaled to best match the secular drift of clocks at Earth's geoid. 
Deep-space radio science and navigation need these time scales in practice: the proper time kept by a clock at a body~X must be tied to the barycentric frame through the same ephemerides and perturber set that propagate the spacecraft and model the signal path, since any mismatch leaks into the estimated range. As range is inferred from a light travel time, $1$\,ns corresponds to about $30$\,cm, so centimetre-level ranging requires TCX$-$TCB at the sub-nanosecond level, where the second post-Newtonian terms become relevant -- besides being there for better consistency.
Rotation modeling is attached to the body-fixed frame and in consequence, in GR, this body-fixed frame has a specific 4-D metric, different from the metric attached to the Barycentric Celestial Reference System (BCRS). Consequently, the time-scale attached to the body-fixed frame in which the Rotation modeling should be described, is different from the TCB or TDB time-scales. However, so far, no rotation modeling except the one for the Earth for which the rotation angles are given relative to TT or TCG, is produced in the natural local time TCX but in TDB.

So far, no general tool exists for computing time-scale for any body that a user can apply to an arbitrary body, with an arbitrary ephemerides, and with a controllable set of perturbers.

We present \tempus, an open-source tool that fills this gap. It numerically integrates the IAU\,2000 time-transformation from an arbitrary planetary ephemerides, and is (i)~\emph{ephemeris-agnostic}, reading INPOP, DE, and EPM files with either TDB or TCB as initial time-scale; (ii)~\emph{body-agnostic}, computing TCX$-$TCB for any target X, or the difference between two body-centered scales such as TCG$-$TCL directly; (iii)~\emph{perturber-aware}, admitting several hundred main-belt asteroids and trans-Neptunian objects; and (iv)~\emph{reproducible}, with a command-line interface (CLI) that specifies a run entirely by its flags and records them in a metadata companion file, a Chebyshev exporter producing standard ASCII files for interpolation, and a Python application programming interface (API) exposing their parts.

The remainder of the paper describes the model and its implementation (Sect.~\ref{sec:model}), with some parameterization caveats discussed in Sect.~\ref{sec:coords}, then validates \tempus against independent solutions (Sect.~\ref{sec:validation}), studies its sensitivity to the input ephemerides and perturbers (Sect.~\ref{sec:perturbers}), describes the Chebyshev polynomial output representation option (Sect.~\ref{sec:cheby}), and then discusses the timing implications for planetary rotation and navigation (Sect.~\ref{sec:applications}) before concluding (Sect.~\ref{sec:conclusions}).

%% file: sections/model.tex

\section{Relativistic Model and Implementation}
\label{sec:model}

\tempus integrates the IAU's post-Newtonian time transformation numerically, evaluating the transformation rate at the center of mass of the considered body X at each step from the positions, velocities, and masses supplied by an arbitrary planetary ephemerides.

\subsection{Post-Newtonian time transformation}
\label{sec:model:pn}

In the framework of the IAU 2000 relativistic reference systems \citep{Soffel2003}, the coordinate time TCX at the center of mass of a body X (located at position $\boldsymbol{r}_X$ with velocity $\boldsymbol{v}_X$ relative to the solar system barycenter) is related to Barycentric Coordinate Time (TCB) by

\begin{equation}
  \frac{d\,\mathrm{TCX}}{d\,\mathrm{TCB}} - 1
  \;=\;
  \frac{\alpha_X}{c^2} + \frac{\beta_X}{c^4}
  \;+\; \OCinq \,,
  \label{eq:dtcx_dtcb}
\end{equation}

\noindent where the $\ODeux$ (first post-Newtonian) term is

\begin{equation}
  \alpha_X = -\frac{1}{2}\,\boldsymbol{v}_X^2 - U_X \,,
  \label{eq:alpha}
\end{equation}

\noindent with the Newtonian gravitational potential at the
position of the center of mass of the body X summed over all other/external bodies $A$:
\begin{equation}
  U_X = \sum_{A \neq X} \frac{G M_A}{r_{AX}} \,,
  \label{eq:UX}
\end{equation}

\noindent and the $\OQuatre$ (second post-Newtonian) correction is

\begin{multline}
  \beta_X = -\frac{1}{8}\,\boldsymbol{v}_X^4 + \frac{1}{2}\,U_X^2 
  + \sum_{A \neq X} \frac{G M_A}{r_{AX}}
  \Bigg[
    4\,\boldsymbol{v}_X \cdot \boldsymbol{v}_A
    - \frac{3}{2}\,\boldsymbol{v}_X^2
    - 2\,\boldsymbol{v}_A^2 \\
    + \frac{1}{2}\,\boldsymbol{a}_A \cdot \boldsymbol{r}_{AX}
    + \frac{1}{2}\left(\frac{\boldsymbol{v}_A \cdot \boldsymbol{r}_{AX}}{r_{AX}}\right)^2
    + \sum_{B \neq A} \frac{G M_B}{r_{BA}}
  \Bigg] \,.
  \label{eq:beta}
\end{multline}

\noindent Here $\boldsymbol{a}_A$ is the barycentric acceleration of body $A$, 
$GM_A$ is its gravitational parameter in the BCRS, 
and $r_{AX}$ is the barycentric distance between body $X$ and $A$.

The $\ODeux$ term $\alpha_X$ is the sum of the specific kinetic and potential energy at the center of mass of body X and produces the familiar secular drift of TCX$-$TCB (for the Earth, $\simeq 1.28$\,ms\,d$^{-1}$) modulated at the orbital period. The $\OQuatre$ term $\beta_X$ collects the $c^{-4}$ cross-terms between the body's motion and the external potential together with the square of the potential itself. For the inner planets $\beta_X/c^4$ contributes a drift around the $10^{-16}$\,s\,s$^{-1}$ level, which accumulates to several hundred nanoseconds per century in TCX$-$TCB, and to $\simeq2.4\,\mu$s per century for Mercury;
so \tempus evaluates the full Eq. \ref{eq:dtcx_dtcb} by default (a \texttt{-{}-pn1} switch is provided to consider only the $\alpha$ contribution for diagnostic comparison). An optional solar oblateness ($J_{2,\odot}$) term is described together with the other perturbing bodies in Sect.~\ref{sec:model:perturbers}.

To fix the free integration constant we can choose to anchor TCX$-$TCB to be zero at the 1977 January~1.0 TAI epoch (JD~$2443144.5003725$ TT), analogous to the Earth case.
In the context of the \tempus integration, this can be achieved by either starting the integration at this date or by applying a shift in post-processing. The latter is recommended, as it allows for arbitrary start epochs and standardised Chebyshev fitting (Sect.~\ref{sec:model:cheby}).
The integrations are run through the CLI command \texttt{tempus integrate} and its options and flags.

\subsection{Perturbing bodies and solar oblateness}
\label{sec:model:perturbers}

The potential sum in Eqs.~(\ref{eq:UX}) and (\ref{eq:beta}) runs over every body with non-negligible mass. The Sun, the eight planetary barycenters, and the Pluto barycenter are always taken from the ephemerides itself and constitute the default external potential set (option \texttt{-{}-ext auto}). Main-belt asteroids (MBAs) and trans-Neptunian objects (TNOs) can be added optionally: their positions are read from a companion trajectory file (\texttt{-{}-asteroid-file}) -- a SPICE small-body kernel such as \texttt{sb441-n373s.bsp} -- while their gravitational masses ($GM$) are taken from the ephemerides constants or supplied through a JPL-style ASCII header (\texttt{-{}-asteroid-gm-header}). Bodies present in the trajectory file but lacking a $GM$ constant or vice versa are skipped. This mechanism supports up to $343$ MBAs and $30$ TNOs in the configurations tested here (Sect.~\ref{sec:perturbers}). The difference in asteroid trajectories between ephemerides have a negligible effect, whereas their mass differences have a noticeable effect, see comparisons in Sect.~\ref{sec:perturbers}.

A further optional perturbation is the Sun's oblateness $J_{2,\odot}$. Beyond the monopole $GM_\odot/r_{\odot X}$ already contained in Eq.~(\ref{eq:UX}), the Sun contributes a quadrupole correction to the $c^{-2}$ potential,
\begin{equation}
  \Delta U_{J_2}
  = -\,\frac{G M_\odot\, J_{2,\odot}\, R_\odot^2}{r_{\odot X}^{3}}\,
  P_2(\sin\phi) \,,
  \label{eq:sunj2}
\end{equation}
with $P_2(x) = \tfrac{1}{2}(3x^2 - 1)$ the second Legendre polynomial,
$r_{\odot X}$ the Sun--target distance, and $\phi$ the heliographic latitude of the target, 
so that $\sin\phi = \hat{\boldsymbol{r}}_{\odot X}\cdot\hat{\boldsymbol{n}}_\odot$
with $\hat{\boldsymbol{n}}_\odot$ the Sun's rotation-pole direction (held fixed at IAU WGCCRE values $\alpha_0 = 286.13\degr$, $\delta_0 = 63.87\degr$; \citealt{Archinal18}). The constants $J_{2,\odot}\approx2.2\times10^{-7}$ and $R_\odot$ are read from the ephemerides header. The term enters the $\ODeux$ potential additively, $U_X\rightarrow U_X + \Delta U_{J_2}$ in Eq.~(\ref{eq:alpha}); it is off by default but may be enabled using the flag \texttt{-{}-sun-j2}. Its effect on the derived time scales is quantified in Sect.~\ref{sec:perturbers:j2}.

\subsection{ephemerides compatibility}
\label{sec:model:ephem}

All ephemerides access goes through CALCEPH \citep[v4.0.5;][]{Gastineau15}, so \tempus reads INPOP binary files (as \texttt{.dat} or \texttt{.tcheb}) and SPICE kernels (\texttt{.bsp} with the associated \texttt{.tpc}/\texttt{.tls}) through a single interface. The timescale of the file is detected from its header. TCB ephemerides are used directly; for a TDB ephemerides \tempus applies the exact linear rescaling of positions, velocities, and $GM$ values by $1/(1-L_B)$, with $L_B = 1.550519768 \times 10^{-8}$ \citep{2010ITN....36....1P}, so that the transformation is always evaluated in TCB-consistent units. Epochs are converted between TT, TDB, and TCB by reading each file's own internal time-ephemerides series rather than an external model, keeping the result self-consistent with the ephemerides in use. All epochs are carried as split (integer\,+\,fractional) Julian dates (JD) to preserve sub-nanosecond resolution, which an ordinary \texttt{float64} JD cannot represent. The ephemerides tested here are INPOP19a (TCB and TDB) \cite{2019NSTIM.109.....V}, INPOP21a (TCB and TDB) \cite{2021NSTIM.110.....F}, DE430 \cite{folkner2014in}, DE440 \cite{Park21}, and EPM2021 \cite{Pitjeva22}.

\subsection{Numerical integration}
\label{sec:model:integrator}

The transformation rate of Eq.~(\ref{eq:dtcx_dtcb}) is integrated with an Adams--Moulton predictor--corrector of order~12 (AM12) by default. Because the integrand is a smooth function of interpolated ephemerides states, a high-order multistep method converges far better, while the total runtime does not increase significantly --- as the quadrature here is a small fraction of the total runtime dominated by ephemerides evaluation calls. 
AM12 is the default after comparison with the trapezoidal rule and lower-order schemes (\texttt{rk4}, \texttt{am4}--\texttt{am10}, all selectable through \texttt{-{}-integrator}). 
The $11$-step start-up phase of AM12 is prepended before the requested start epoch, such that the reported output series is free of start-up transients. 
All results here use an integration step size of $0.75$\,d ($18$\,h), matching the internal sampling rate of the ephemerides like INPOP.

\subsection{Chebyshev polynomial output}
\label{sec:model:cheby}
For downstream use the integrated samples can be fitted, by the \texttt{tempus chebyshev} subcommand, into a piecewise Chebyshev series and written as an ASCII coefficient file (\texttt{.asc}) in the same tabular layout as the series distributed by INPOP and \citet{Klioner}, which \tempus also reads directly for comparison. The subcommand can also write the series in either of the two time-argument parameterisations distributed with the TU~Dresden time solutions (see Sect.~\ref{sec:coords}). The fitting algorithm, its tuning options, and the representation and evaluator accuracy are described in Sect.~\ref{sec:cheby}. To keep the fitting and physical error budgets separate, all comparisons in Sect.~\ref{sec:validation} use the direct \tempus integration results directly, without Chebyshev fitting.

%% file: sections/coords.tex

\section{Coordinate and parameterisation choices}
\label{sec:coords}

Two options in \tempus change not the accuracy but the very quantity that is computed, and both are easy to misuse. We highlight them here because a user who overlooks this would incur errors and inconsistencies far larger than the sub-nanosecond model accuracy established in Sect.~\ref{sec:validation}.

\subsection{Differences between two body-centered scales}
\label{sec:coords:center}

By default the integration is centered on the solar-system barycenter (\texttt{-{}-center SSB}), giving TCX$-$TCB at the body X center of mass. Setting \texttt{-{}-center} to a body and \texttt{-{}-target} to another --- for example \texttt{-{}-center EARTH\_CENTER -{}-target MOON} --- makes \tempus run two barycentric integrations and subtract them, returning the difference between two time-scales at the centers of mass of the two bodies, here TCG$-$TCL, as a function of TDB or TCB (depending on what timescale the underlying ephemerides is parameterised with, typically TDB). This subtraction is exact \emph{as a function of TDB or TCB}. 

One must be aware though, that this difference is only valid in the simultaneity of the BCRS. To move to a different notion of simultaneity, the user must add a kinematic boundary term. In the case of TCL, it would be $\mathbf{v}_E\!\cdot\!\mathbf{r}_{LE}/c^2$, that moves to a geocentric-simultaneity description. This term has an amplitude of $\sim125$\,\textmu s oscillating at the lunar month, and is only a function of the ephemerides built from the same state vectors as the \tempus integration itself.

\tempus deliberately leaves this term out of the default output and provides the function \texttt{tempus.model.boost\_term} to add it when a particular simultaneity convention is required.
The two descriptions are different functions of time; consequently a naive subtraction of two TCX$-$TCB series (staying in the simultaneity of the BCRS), compared against a reference such as the lunar solution in \cite{Kopeikin_2024} (which carries a GCRS notion of simultaneity), will disagree by this $\sim125$\,\textmu s term. This is expected behaviour, not an error.

What has been described so far has exclusively been center-of-mass quantities. A real clock, at the surface or in orbit, needs one further term $c^{-2}(-v_\mathrm{clock}^{2}/2 - W_X(\mathbf{x_\mathrm{clock}}))$, which unlike the boundary term above is not a function of the ephemerides: it requires a gravity-field model of body $X$ and a clock location/trajectory, and thus lies outside the scope of \tempus and is a users responsibility. This term dominates any clock comparison, and is applied on top of (each) TCX$-$TCB.

\subsection{Choice of time argument}
\label{sec:coords:argument}

The offset TCX$-$TCB tabulated \emph{as a function of TCB/TDB} and the same offset tabulated \emph{as a function of the body's own time at its center of mass} TCX are two different functions. They differ by a secular term $F\cdot\Delta \approx L_X^2\,\tau$ that grows linearly with elapsed time $\tau$ and quadratically with the body's rate constant $L_X$: about $0.69$\,\textmu s\,cy$^{-1}$ for TCG and $0.30$\,\textmu s\,cy$^{-1}$ for TCM \citep{Klioner2026}. Evaluating a series built against one argument at a numerically equal value of the other therefore introduces an error of exactly this magnitude --- microseconds per century, orders of magnitude above the model accuracy. Any user of a time-ephemerides product must know which argument parameterises it.

\tempus makes both available. The integration and its native ASCII export are parameterised by the ephemerides time argument (TCB for a TCB ephemerides, TDB for a TDB one); \texttt{tempus chebyshev -{}-also-tcx} additionally writes the offset parameterised by the body's own coordinate time TCX. Because every integrated sample is a single event carrying both time labels, this re-parameterisation is an exact per-event relabelling of the output (\texttt{tempus.model.relabel\_to\_tcx}) rather than a re-integration, with \texttt{-{}-anchor} aligning the relabelled axis to the 1977 origin. This re-labeling applied either to a \tempus run, or separately to the TCB-parameterised ASCII file distributed by \citet{Klioner}, reproduces Klioner's independently-integrated, TCX-parameterised (``inverse'') product at the picosecond level.

%% file: sections/validation.tex

\section{Validation}
\label{sec:validation}

We validate \tempus by comparing its output against other solutions: the independent TU~Dresden solution \citep{Klioner}, the TCG$-$TCB series stored inside the ephemerides files themselves INPOP19a and INPOP21a, and the LTE440 lunar time solution \citep{Lu2025}. As will be seen, these comparisons show differences far below the accuracy of the modern atomic clocks, and validate our implementation with \tempus.

A complementary question, how much the result depends on the input ephemerides and on the considered dynamical modeling, is a sensitivity study that we defer to Sect.~\ref{sec:perturbers}. 

In Tables \ref{tab:klioner} and \ref{tab:validation} each comparison row reports the difference of two solutions. All series are sampled on a common $0.75$\,d ($18$\,h) grid. Differences of solutions consist of a linear \emph{drift} (the fitted slope) and remaining residual after detrending, which is described via its \emph{peak} amplitude. Drifts are quoted in fractional-rate units (s\,s$^{-1}$) so they can be compared directly with clock-accuracy figures; $1$\,ns\,cy$^{-1}=3.169\times10^{-19}$\,s\,s$^{-1}$. 

As a baseline consistency check, running on the TDB and TCB releases of INPOP19a, as shown in the first row of Table~\ref{tab:validation}, at matched physical epochs yields series that agree to $\sim10^{-23}$\,s\,s$^{-1}$ (sub-picosecond over the full 200-yr span), confirming that the $1/(1-L_B)$ rescaling of positions and $GM$ values (Sect.~\ref{sec:model:ephem}) is applied correctly and that \tempus is genuinely agnostic to the timescale of its input.


\begin{table}
\caption{\tempus against the independent TU~Dresden solution
\citep{Klioner} for every body it publishes, from the files at
\url{https://gaia.geo.tu-dresden.de/TimeEphemerides/}. Both solutions here
use INPOP19a with as perturbers the Sun and major planets only over the same $200$\,yr span, so each row isolates the transformation itself. The first pair of columns gives the secular rate of the quantity compared, on which both solutions agree to all nine digits computed; the second pair gives the differences in the same fractional-rate units. For the Earth's case, also see Fig.~\ref{fig:tcg_klioner}.
The \cite{Klioner} series are stitched from forward and backward integrations meeting at the 1977 origin; the resulting step, at most $5.4$\,ps, is removed here before linear fitting.}
\label{tab:klioner}
\centering
\setlength{\tabcolsep}{4pt}
\begin{tabular}{llrrr}
\hline\hline
      &  &  & \multicolumn{2}{c}{TEMPUS vs Klioner} \\
\cline{4-5}
      & \multicolumn{2}{c}{TCB$-$TCX drift} & Drift & Peak \\ 
Body  & [ms\,d$^{-1}$] & [$10^{-10}$\,s/s] & [$10^{-21}$\,s/s] & [ps] \\
\hline
Sun     & $0.000192$ & $0.022$   & $8.71$ & $0.4$  \\
Mercury & $3.304862$ & $382.507$ & $4.07$ & $13.6$ \\
Venus   & $1.768717$ & $204.713$ & $2.19$ & $24.8$ \\
Earth   & $1.279434$ & $148.083$ & $5.04$ & $5.9$  \\
Moon    & $1.280911$ & $148.254$ & $4.30$ & $6.9$  \\
Mars    & $0.839753$ & $97.194$  & $3.37$ & $4.9$  \\
Jupiter & $0.245831$ & $28.453$  & $7.28$ & $0.6$  \\
Saturn  & $0.134269$ & $15.540$  & $1.65$ & $1.1$  \\
Uranus  & $0.066791$ & $7.730$   & $0.62$ & $0.2$  \\
Neptune & $0.042585$ & $4.929$   & $0.38$ & $0.1$  \\
Pluto   & $0.037345$ & $4.322$   & $0.33$ & $0.2$  \\
\hline
\end{tabular}
\end{table}

\begin{table*}
\caption{Validation of \tempus against the other available TCX$-$TCB
solutions besides \cite{Klioner} solution (see Table \ref{tab:klioner}). Each solution is done on an 200 year interval. The ephemerides each solution uses is given in parentheses, and ``INPOP'' denotes the solution stored inside the INPOP file (queried through CALCEPH), available for the Earth alone. ``Drift'' is the slope of a linear fit to the difference $\Delta(\text{TCX}-\text{TCB})$; ``peak'' is the maximum amplitude of the detrended difference. All rows are evaluated at the same events on one TDB grid. The last row crosses ephemerides (\tempus on INPOP19a against a DE440 product) and is discussed in Sect.~\ref{sec:vali:lte440}.}
\label{tab:validation}
\centering
\begin{tabular}{lllrrr}
\hline\hline
Body & Comparison & \tempus Model  & Drift [$10^{-18}$\,s/s] & Peak [ps] \\
\hline
\multicolumn{5}{l}{\emph{Internal consistency of \tempus}} \\
Earth & \tempus{}(INPOP19a\,TDB) $-$ \tempus{}(INPOP19a\,TCB)  & planets only  & $<0.0001$ & $<0.1$ \\
\hline
\multicolumn{5}{l}{\emph{Against the INPOP solutions}}\\
Earth & \tempus{}(INPOP19a) $-$ INPOP(19a)  & with 353 small bodies & $-0.006$ & $11$ \\
Earth & \tempus{}(INPOP21a) $-$ INPOP(21a)  & with 353 small bodies & $17.596$ & $254$  \\
Earth & \tempus{}(INPOP19a) $-$ INPOP(19a)  & planets only    & $6.188$ & $173$  \\
Earth & \tempus{}(INPOP21a) $-$ INPOP(21a)  & planets only    & $24.162$ & $331$  \\
\hline
\multicolumn{5}{l}{\emph{Against an independent third-party solution}}\\
Moon  & \tempus{}(DE440) $-$ LTE440(DE440)    & with 373 small bodies & $16.404$ & $43$  \\
Moon  & \tempus{}(INPOP19a) $-$ LTE440(DE440) & with 353 small bodies & $17.710$ & $257$  \\
\hline
\end{tabular}
\end{table*}



\begin{figure}
  \centering
  \includegraphics[width=\columnwidth]{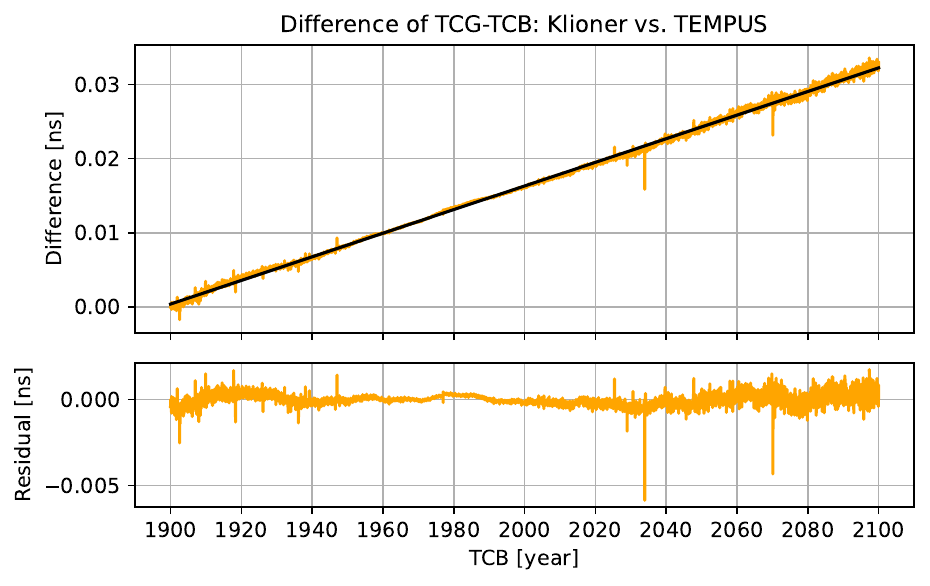}
  \caption{Independent validation of the TCG$-$TCB implementation: \tempus minus the TU~Dresden (Klioner) solution over 200\,yr, both computed from INPOP19a. Top: the difference; the fitted slope
  in black has a drift of $5.04\times10^{-21}$\,s\,s$^{-1}$ ($0.016$\,ns\,cy$^{-1}$). Bottom: the detrended difference stays below $6$\,ps. This confirms our implementation of Eq.~(\ref{eq:dtcx_dtcb}).}
  \label{fig:tcg_klioner}
\end{figure}


\subsection{Comparison with the Klioner solution.}
The strongest test available is against a solution produced by different authors using different code from the exact same input. 
The TU~Dresden time solutions \citep{Klioner} provide exactly that for eleven bodies. Table~\ref{tab:klioner} compares every one of them against a comparable evaluation using \tempus. Both solutions agree on the secular rate of TCX$-$TCB to all nine digits computed (six given in the table); the disagreement is confined to a remaining relative drift between $3.3\times10^{-22}$ and $8.7\times10^{-21}$\,s\,s$^{-1}$ ($1$--$30$\,ps\,cy$^{-1}$) with detrended peak values below $10$\,ps, except for Mercury and Venus. Also see Fig.~\ref{fig:tcg_klioner} for a detailed Earth's comparison plot specifically.

The detrended difference (in the table proxied by the peak column) is not the same for every body: it seems to track orbital speed rather than the magnitude of the offset itself, which is largest for Mercury. The fastest worldlines are where two independent solutions have the most opportunity to differ. The peak column is nevertheless an outlier-dominated statistic: looking at the data directly, Venus shows the largest peak ($24.8$\,ps as seen in Table~\ref{tab:klioner}) on an RMS of about $0.6$\,ps, while Mercury's RMS is twice as large, so the ordering by RMS follows orbital speed more faithfully than the ordering by peak.

The two solutions are numerically independent in a strong sense: While \tempus precomputes the rate on a fixed grid and integrates it with a fixed-order multistep scheme, \cite{Klioner} integrates the differential equation directly with the adaptive-order extrapolation code ODEX.
Our excellent agreement with the very small observed relative drifts, therefore already validates the implementation of Eq. (\ref{eq:dtcx_dtcb})'s time transformation in \tempus.

\subsection{Comparison with the time solution provided by INPOP ephemerides.}
INPOP distributes its own TCG$-$TCB series together with the ephemerides file, which allows a second, quite different check: against the time scale that the ephemerides publishers itself derived while constructing the ephemerides. 
But the two dynamical models are not necessarily identical by construction. \tempus evaluates the potential from the bodies the user supplies; INPOP built its internal series from the full perturber set of its own fit. Any mismatch in that body list, or lack of an idealized internal modeling of say Kuiper belt rings, is a physical signal, not a numerical error, and the comparison is best read as a measurement of how completely the dynamical model has been reproduced.

\begin{figure}
  \centering
  \includegraphics[width=\columnwidth]{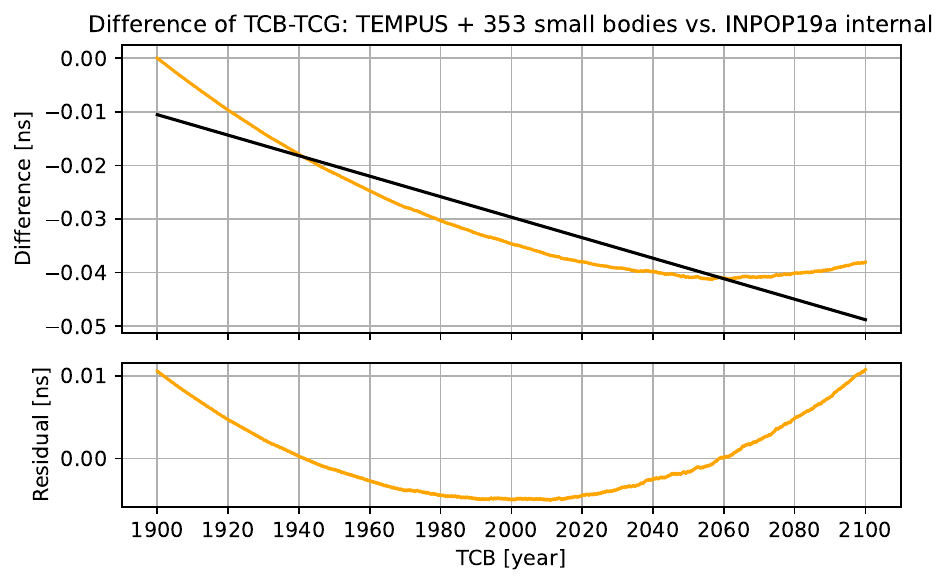}
  \caption{Differences of TCG$-$TCB via \tempus and the independent solution distributed by the INPOP19a release. Sun, Planets, Pluto and the standard INPOP19a 353 small bodies are inlcuded in the calculation by \tempus. Linear fit in black, that shows a relative drift of $-6\times10^{-21}$\,s\,s$^{-1}$ ($-19$\,ps\,cy$^{-1}$) with a   residual max peak of $11$\,ps. Agreement is at the same order of magnitude, as in Fig.~\ref{fig:tcg_klioner}}
  \label{fig:tcg_353ast}
\end{figure}

This interpretation is supported by Table~\ref{tab:validation}. With a planets-only potential the difference against INPOP19a drifts at $6.19\times10^{-18}$\,s\,s$^{-1}$ ($19.5$\,ns\,cy$^{-1}$) over $200$\,years, and the \cite{Klioner} solution differs from the same INPOP internal series by an identical slope and shape --- so the drift belongs to the missing bodies, not to either implementation. Adding the $353$ standard objects of INPOP19a (343 MBAs and 10 TNOs) reduces the drift by a factor $1021$, to $-6\times10^{-21}$\,s\,s$^{-1}$ ($-0.019$\,ns\,cy$^{-1}$), with a residual of $11$\,ps (see Fig.~\ref{fig:tcg_353ast}), closing the consitency gap so the same order-of-magnitude agreement with \cite{Klioner}, as seen before in Fig.~\ref{fig:tcg_klioner}. 


The same exercise on INPOP21a behaves differently, and the contrast is the more interesting result, which probes how INPOP21a differs from INPOP19a regarding Kuiper belt objects. Its planets-only difference is already a factor $4$ larger ($2.42\times10^{-17}$\,s\,s$^{-1}$) than for \tempus-INPOP19a planet-only run, and adding the $353$ small bodies we had available for that release improves it by only a factor $1.4$, to $1.76\times10^{-17}$\,s\,s$^{-1}$. Supplying the known 353 point masses, in other words, does not close the gap against INPOP21a the way it does against INPOP19a. The higher relative drift seems to be caused by our \tempus run not accounting for $500$ individually integrated objects in INPOP21a \citep{Fienga21}, that amount to a quoted total mass of $0.0259$\,M$_\oplus$ between $39.3$ and $47.6$\,AU. That mass contributes a near-constant potential at the Earth of about $U/c^2 = 1.77\times10^{-17}$\,s\,s$^{-1}$ -- matching the above observed differential drift w.r.t. INPOP21a to better than $1\%$.

\subsection{Comparison with LTE440.}\label{sec:vali:lte440}
The lunar time solution LTE440 \citep{Lu2025} is an independent product built on DE440, with the claim it completely reproduces DE440's dynamical model. Run on that same DE440 ephemerides, \tempus reproduces it to $1.64\times10^{-17}$\,s\,s$^{-1}$ ($52$\,ns\,cy$^{-1}$) with a detrended peak of $0.043$\,ns. This difference is not a floor: it falls from $2.38$ to $1.64\times10^{-17}$\,s\,s$^{-1}$ as the $343$ main-belt asteroids and the $30$ individual TNOs are added versus a planets-only run, with the detrended peak dropping by a  factor $4.5$. It therefore still measures the remaining difference in dynamical model rather than the transformation itself.

Repeating the comparison with \tempus using INPOP19a, barely changes the difference: $2.39\times10^{-17}$\,s\,s$^{-1}$ with planets-only and
$1.77\times10^{-17}$\,s\,s$^{-1}$ with INPOP19a's 353 small bodies, with detrended peaks of $0.37$ and $0.26$\,ns. What Table~\ref{tab:validation}'s last row measures is the INPOP19a-versus-DE440 difference in the lunar ephemerides dynamical model, and it is included here only to show how completely the choice of input ephemerides can dominate a comparison --- the question taken up systematically in the following section.


%% file: sections/perturbers.tex

\section{Sensitivity study}
\label{sec:perturbers}

The transformation rate of Eq.~(\ref{eq:dtcx_dtcb}) is built entirely from the masses, positions, and velocities of the bodies that make up the solar system: the potential $U_X$ (Eq.~\ref{eq:UX}) and the $c^{-4}$ cross-terms (Eq.~\ref{eq:beta}) sum over every perturber with a provided $GM$ and trajectory. The computed time scale is therefore only as good as its underlying dynamical model, and crucially that model must be kept \emph{consistent} with the ephemerides being used: \tempus should evaluate the potential from the same bodies, with the same masses and trajectories, that were used to construct the ephemerides in the first place. A mismatch between the two body sets does not degrade the numerical accuracy of the transformation; it injects a physical signal --- the gravitational effect of the bodies present in one model but absent in the other.

Two levers control this dynamical model, and we vary them independently. The choice of planetary ephemerides fixes the masses and orbits of the Sun and planets in a mutually consistent way (Sect.~\ref{sec:perturbers:ephem}). On top of a given ephemerides, the optional small-body perturbers --- main-belt asteroids (MBAs) and trans-Neptunian/Kuiper-belt objects (TNOs or KBOs) --- and the solar oblateness can be switched on and off (Sects.~\ref{sec:perturbers:mba} and \ref{sec:perturbers:j2}), isolating each contribution while the planetary motion is held fixed. We quantify the sensitivity of TCX$-$TCB to each.

One limitation applies throughout. Modern ephemerides may model parts of the Kuiper belt as an ideal ring of evenly spaced and equal mass point masses on fixed Keplerian orbits, rather than an extended set of actual small bodies. Such a treatment is not yet supported by \tempus.

\begin{table*}[!t]
\caption{Sensitivity of TCX$-$TCB, measured as the difference of two \tempus runs identical except one property, over a common $200$\,yr span centered around J2000. Each row names both runs, so a row differing only in the ``Model'' columns isolates a physical term and one differing only in the ``ephemerides'' columns isolates the ephemerides. ``Drift'' is the slope of a linear fit to the difference, in fractional-rate units; ``Peak'' is the largest deviation once that linear slope is removed from the difference. Drifts are magnitudes. In block (c) ``own asteroids'' means each ephemerides's native small-body set (INPOP19a/21a: 353; DE430: 343; EPM2021: 244).}
\label{tab:ephemerides}
\centering
\begin{tabular}{lcccrrr}
\hline\hline
 & \multicolumn{2}{c}{\tempus{}$(x)$} & \multicolumn{2}{c}{\tempus{}$(y)$} & Drift & Peak \\
\cline{2-3}\cline{4-5}
Body & ephemerides & Model & ephemerides & Model & [$10^{-18}$\,s/s] &  [ps] \\
\hline
\multicolumn{7}{l}{\emph{(a) The hierarchy of model terms, for one body on one ephemerides}}\\
Earth & INPOP19a & Eq.~\ref{eq:dtcx_dtcb} with only $\alpha$ term    & INPOP19a & Full Eq. \ref{eq:dtcx_dtcb}  & $109.64$ & $34$         \\
Earth & INPOP19a & planets       & INPOP19a & $+$ 353 bodies & $6.19$ & $160$       \\
Earth & INPOP19a & planets       & INPOP19a & $+$ Sun $J_2$  & $0.02$  & $1$      \\
\hline
\multicolumn{7}{l}{\emph{(b) How many small bodies are needed, and what the distant population adds}}\\
Earth & INPOP19a & planets       & INPOP19a & $+$ 5 MBAs     & $2.76$ & $13$  \\
Earth & INPOP19a & planets       & INPOP19a & $+$ 343 MBAs   & $4.34$ & $13$  \\
Earth & INPOP19a & $+$ 343 MBAs  & INPOP19a & $+$ 10 TNOs    & $1.85$ & $161$  \\
Mars  & INPOP19a & planets       & INPOP19a & $+$ 5 MBAs     & $2.92$ & $71$  \\
Mars  & INPOP19a & planets       & INPOP19a & $+$ 343 MBAs   & $4.58$ & $72$  \\
Mars  & INPOP19a & $+$ 343 MBAs  & INPOP19a & $+$ 10 TNOs    & $1.85$ & $157$  \\
Moon  & DE440    & planets       & DE440    & $+$ 5 MBAs     & $2.74$ & $12$  \\
Moon  & DE440    & planets       & DE440    & $+$ 343 MBAs   & $4.71$ & $13$      \\
Moon  & DE440    & $+$ 343 MBAs  & DE440    & $+$ 30 KBOs    & $2.64$ & $187$       \\
\hline
\multicolumn{7}{l}{\emph{(c) ephemerides changed, each with its own small-body set}}\\
Mars  & INPOP19a & own asteroids & INPOP21a & own asteroids  & $0.36$ & $608$ \\
Mars  & INPOP21a & own asteroids & DE430    & own asteroids  & $2.10$ & $1421$      \\
Mars  & INPOP21a & own asteroids & EPM2021  & own asteroids  & $2.44$ & $536$      \\

Mars  & DE430    & own asteroids & EPM2021  & own asteroids  & $0.34$  & $1676$      \\
\hline
\end{tabular}
\end{table*}

All \tempus runs and comparisons relevant for this section are found in Table~\ref{tab:ephemerides}. Section (a) of the Table gives an overview of the expected incurred relative drift values, if enabling or disabling the full usage of Eq. \ref{eq:dtcx_dtcb}, all small bodies, or the Sun J$_2$ term. The biggest relevant effect, as stated in the beginning is by the contribution of the $\beta$ term in Eq. \ref{eq:dtcx_dtcb}, on a scale of $10^{-16}$\,s\,s$^{-1}$, followed by the effect of small bodies on the $10^{-18}$\,s\,s$^{-1}$ scale, and with the least effect by the solar oblateness on the scale of $10^{-20}$\,s\,s$^{-1}$.

\subsection{Sensitivity to planetary ephemerides}
\label{sec:perturbers:ephem}

Because \tempus post-processes an existing ephemerides, it inherits the differences between input ephemerides.  Here we quantify how much the choice of input matters. 
Block~(c) of Table~\ref{tab:ephemerides} does this for Mars, the body whose time scale is most exposed to the question: it orbits closest to the main belt, so it is most susceptive to asteroid perturbations. Each ephemerides are taken as a user should process it, with its own native small-body set, so a row measures the total change on switching ephemerides input.

Within the INPOP family the step from INPOP19a to INPOP21a moves TCM$-$TCB by $3.6\times10^{-19}$\,s\,s$^{-1}$ ($1.1$\,ns\,cy$^{-1}$). Across independent families (INPOP, the JPL DE, and the IAA EPM series, each built from overlapping but not identical observation sets and dynamical models) the differences are roughly six times larger, $2.10\times10^{-18}$ between INPOP21a and DE430 and $2.44\times10^{-18}$ between INPOP21a and EPM2021. DE430 and EPM2021 agree with each other far better than either agrees with INPOP, at $3.4\times10^{-19}$. This can be explained by the differences in the dynamical modelings of these ephemerides (see \cite{fienga2024} for a review).
The same comparison for the Moon, computed for the ILuRS2026 lunar products over the $1970$--$2052$ span, gives a consistent picture. As given by \cite{ilurs}, the differences estimated over the ILuRS2026 interval are $1.73\times10^{-18}$ between INPOP21a and DE430, $1.89\times10^{-18}$ between INPOP21a and EPM2021, and $1.61\times10^{-19}$ between DE430 and EPM2021. The ordering is the same as for Mars and the drifts agree with the Martian values to within about $30\%$, so the bound on uncertainty induced by planetary and lunar ephemerides is not peculiar to one target. 

Overall the choice of input ephemerides affects TCX$-$TCB at the few\,$\times10^{-18}$\,s\,s$^{-1}$ level and below, with nanosecond-level periodic structure.

\subsection{Sensitivity to asteroid and Kuiper-belt number}
\label{sec:perturbers:mba}

Holding the ephemerides fixed, we now switch the small-body perturbers on and off (Section~(b) of Table~\ref{tab:ephemerides}). Every user-provided body with resolvable mass and position state contributes to the potential $U_X$ (Eq.~\ref{eq:UX}). 

The most massive main-belt asteroid, Ceres, is comparable to a small planetary moon; together with Vesta, Pallas, Iris and Bamberga it dominates the belt's mass and gravity. We indeed find that these five asteroids (which also enter INPOP's equations of motion) account for the bulk of the belt: they shift TCX$-$TCB by $2.8\times10^{-18}$\,s\,s$^{-1}$ for the Earth and $2.9\times10^{-18}$ for Mars, against $4.3$ and $4.6\times10^{-18}$ for the complete $343$-object main-belt set. These five bodies thus account for about $60\%$ of the belt's effect. The same holds on DE440 for the Moon.

The distant population enters the potential in the same way but from some fifteen times further out, and its effect is correspondingly smaller without being negligible. Individually these objects are far more massive --- Eris has $GM\approx1100$\,km$^3$\,s$^{-2}$  against Ceres $GM\approx62$ km$^3$\,s$^{-2}$ --- so their contributions to $GM/r$ are comparable body for body, and only their small number keeps the total down. Added on top of the complete main belt, the $10$ distant objects distributed with INPOP19a change TCX$-$TCB by $1.9\times10^{-18}$\,s\,s$^{-1}$ for both the Earth and Mars, about $40\%$ of what the entire main belt contributes; the $30$ KBOs of DE440 give $2.6\times10^{-18}$ for the Moon, a larger share on account of their greater number.

Runs that include the distant population also show much larger peaks --- $157$--$187$\,ps against $13$\,ps (Earth, Moon) and $72$\,ps (Mars) for the main-belt-only runs. We attribute this to orbital periods involved rather than from any modeling deficiency: objects beyond $40$\,AU complete well under one revolution in $200$\,yr, so their contribution to $U_X$ is a slowly curving drift that a linear fit cannot absorb, and it survives almost entirely into the residual. Main-belt asteroids cycle roughly forty times over the same window, so their variation averages down to a few tens of picoseconds at most.

\subsection{Effect of the solar oblateness}
\label{sec:perturbers:j2}

The Sun's quadrupole moment $J_{2,\odot}$ enters the $c^{-2}$ potential (Sect.~\ref{sec:model:pn}). Its effect on the planetary time scales is negligible: for the Earth (TCG) on INPOP19a it contributes a drift of $2.1\times10^{-20}$\,s\,s$^{-1}$, with a residual within $\pm0.6$\,ps over 200\,years. This is four orders of magnitude below the $\beta$ term of Eq.~(\ref{eq:dtcx_dtcb}) and two under effect of small bodies. 
It is off by default, however a user may enable it with \texttt{-{}-sun-j2}.

%% file: sections/chebyshev.tex

\section{Chebyshev representation}
\label{sec:cheby}

The integrated samples are the primary output of \tempus, but the form in which TCX$-$TCB is distributed and consumed downstream is a piecewise Chebyshev polynomial series in a plain-text ASCII coefficient file (Sect.~\ref{sec:model:cheby}). 
This section describes the fitting algorithm, the options that control it, and the accuracy of the resulting Chebyshev product, in both the \emph{representation} error of approximating the integrated series by these polynomials, and the \emph{evaluation} error of reading those coefficients back.

\subsection{Fitting algorithm}
\label{sec:cheby:algo}

The span is divided into multiple fixed-length segments, called \emph{granules} (default $4$\,d), each with a fitted Chebyshev series of fixed degree. 

Within a granule the fit is a collocation rather than a least-squares problem: $n{+}1$ collocation nodes are placed on $[-1,1]$, the \tempus samples are interpolated to those nodes, and the Chebyshev coefficients follow from a single small linear solve $\mathbf{c}=\mathsf{M}^{-1}\mathbf{f}$, where $\mathsf{M}$ is the Vandermonde matrix of the basis at the nodes. Because $\mathsf{M}^{-1}$ depends only on the degree and node distribution, it is computed once and cached, so fitting each granule costs one matrix--vector product; the granules are independent and fitted in parallel. Evaluation uses the Clenshaw recurrence, and all times are carried relative to the granule start to avoid cancellation against the large ($\sim2.4\times10^{6}$) Julian-date base.

The mentioned interpolation of the \tempus integration samples to the nodes is done with a local \emph{stencil} of \texttt{-{}-npoints} neighbouring samples (default $10$). If the integration rate $d\,\mathrm{TCX}/d\,\mathrm{TCB}$ is exported alongside the values (\texttt{integrate -{}-out-rate}), a Hermite interpolant using both value and derivative may be used; otherwise a Lagrange interpolant on the values alone is used. Three node distributions are available (\texttt{-{}-dist}): Gauss--Lobatto with endpoints (default), the interior Chebyshev--Gauss roots, and a $C^1$-continuous Lobatto variant that additionally matches the derivative at the granule boundaries --- enforcing continuity of both value and rate across granules --- which requires the rate column.

\subsection{Tuning the fit}
\label{sec:cheby:tuning}

The representation error is governed by the polynomial degree (\texttt{-{}-order}) and the granule length (\texttt{-{}-granule-len}): raising the degree or shortening the granule both reduce the residual, trading file size for accuracy. \tempus gates the fit on two independent quality checks. The first is a truncation estimate from the decay of the trailing Chebyshev coefficients, which flags a granule whose series has not converged. The second is an explicit residual budget (\texttt{-{}-max-err}): the maximum $|\text{fit}-\text{samples}|$ over the whole span must stay below a user-set threshold. The budget is the more reliable gate near machine precision, where the coefficient tail flattens into the float64 noise floor and the truncation test can report non-convergence for an otherwise excellent fit. When a gate fails, \texttt{-{}-auto} retries with progressively higher degree and halved granule length until the budget is met. Finally, \texttt{-{}-anchor} zeros the series at a chosen epoch (bare \texttt{-{}-anchor} uses the 1977 origin) by shifting only the constant coefficient of each granule, as long as the \tempus integration run covers that chosen anchor epoch.

\subsection{Representation and evaluator accuracy}
\label{sec:cheby:validation}

The representation error is not a fixed property of the method but a setting: it is bounded by the residual budget \texttt{-{}-max-err}, which the fit is gated on (Sect.~\ref{sec:cheby:tuning}), and can be pushed down by raising the degree or shortening the granule at the cost of file size. A user can therefore choose it to sit as far below the physical error budget of Sect.~\ref{sec:validation} as the application requires.

The evaluator is validated independently against CALCEPH. An INPOP time-ephemerides ASCII is an export of the Chebyshev coefficients stored inside the binary ephemerides, so reading it with the \tempus evaluator and querying the same quantity through CALCEPH exercises two implementations -- our Clenshaw recurrence  implemented as part of \tempus and CALCEPH's own recurrence -- on identical coefficients. Over $10^{5}$ random epochs spanning all granules of the INPOP19a TCG$-$TCB file the two agree to at most one unit in the last place ($\sim4\times10^{-16}$\,s), with $99.99\%$ of epochs bit-identical, confirming that the evaluation adds no error above float64 rounding: reading a distributed series back costs nothing against the model accuracy.

%% file: sections/applications.tex

\section{Applications}
\label{sec:applications}

A body's coordinate time has two kinds of use, and we give one example of each. The first is a modeling question independent of any mission: rotation models are currently referred to TDB while the rotation itself belongs in the body's own time scale, and Sect.~\ref{sec:app:rotation} quantifies the difference. The second is operational: the radio navigation systems being designed for the Moon, Mars need a body time that is traceable to Earth at the nanosecond level, and Sect.~\ref{sec:app:missions} asks whether the transformation meets that. \tempus supplies the transformation only; it does not replace the orbit determination, tracking or rotation models that consume it.

\subsection{Relativistic timescales for planetary rotation}
\label{sec:app:rotation}

\begin{figure*}
\centering
\includegraphics[width=0.9\hsize]{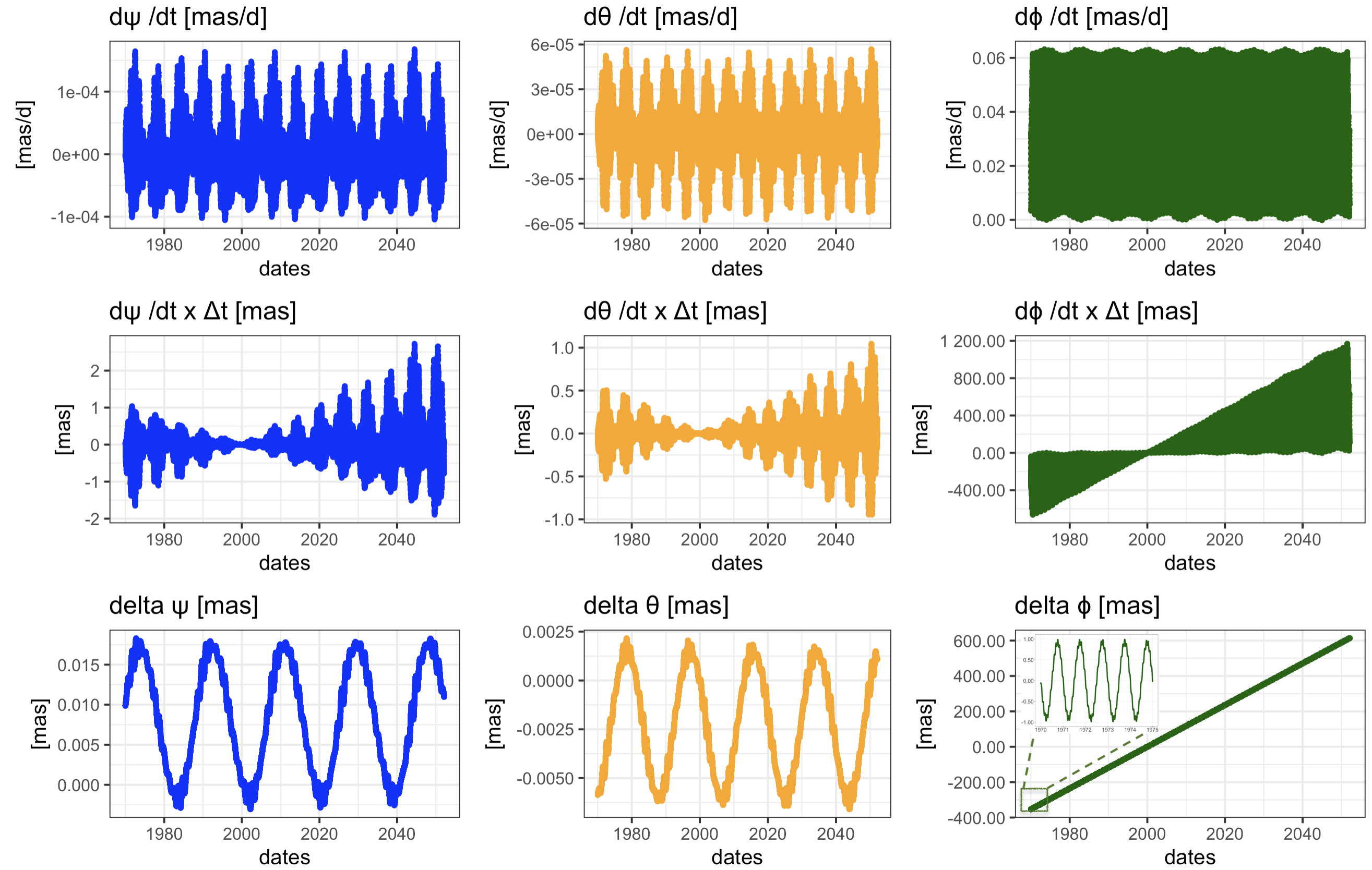}
\caption{Top: Differences between the TCL and TDB time derivatives of the orientation angles $\psi$, $\theta$, $\phi$ (mas/d), from Eq.~(\ref{eq:PATCL}). Middle: rate differences multiplied by the time elapsed since J2000 $\Delta t$, an upper bound on the accumulated angle differences (mas). Bottom: the angle differences obtained by integrating the top-row series from J2000 (mas). The inset panel gives the periodic signals after the removal of the drift for $\phi$. The Euler angles are $\psi$, the precession angle, $\theta$, the nutation angle, and $\phi$, the free rotation angle.}
\label{fig:PATCL}
\end{figure*}

A body's rotation must be modeled in the frame attached to it, known as the body-fixed frame. In general relativity, this frame is defined by an origin at the body's center of mass, a local metric determined by the body's own gravitational field, and an orientation that evolves under the external gravitational torque from other bodies. \citet{Klioner2003} proposed in 2003 the only framework that describes orbital and rotational motion for the Earth self-consistently within general relativity. For the Moon, orbit and rotation are integrated together numerically \citep{viswanathan2018,pavlov2016}: the orbit follows the full parametrised post-Newtonian (PPN) Einstein-Infeld-Hoffmann (EIH) equations \citep{Soffel2003}, but rotation uses the Newtonian Liouville equations, with added relativistic terms, as in state-of-the-art ephemerides \citep{Park21, 2021NSTIM.110.....F, Pitjeva2014}. However, these Newtonian Liouville equations are solved not in the Moon's own time-scale, TCL, but in TDB, the common time-scale used to solve the EIH equations for all solar-system bodies. Dating the Moon's rotation in barycentric rather than local time — this time-rotation coupling — remains standard practice in current rotation models.
It is then possible to do a conversion from the orientation Euler angles (PA) given in TDB as described in \cite{Rambaux2026} to PA in TCL in consider a change of variable such as:
\begin{equation}
    \frac{d PA}{d TCL} = \frac{d PA}{d TDB} \times \frac{d TDB}{d TCB}\times \frac{d TCB}{d TCL}
\label{eq:PATCL}
\end{equation}
With angles not being affected by time-scales, only the time derivatives of the orientation angles PA have to be transformed following Eq.~(\ref{eq:PATCL}). 

The differences between the TCL and TDB time derivatives of the orientation angles are plotted in Fig.~\ref{fig:PATCL} top panels, in the middle panels the rate differences multiplied by the time elapsed since J2000, and their integrals in the bottom panels. The most significant effect is on the time variation of the free rotation angle $\phi$, with a maximum amplitude of about 0.063 milliarcseconds (mas) per
day.

On the top panels of Fig.~\ref{fig:PATCL}, $\dot{\psi}$ and $\dot{\theta}$ oscillate symmetrically about zero, remaining bounded with no drift in the envelope center. $\dot{\phi}$, by contrast, carries an offset of about $0.03$\,mas/day and a periodic modulation superimposed on a non-zero mean, which is the signature of a secular rate combined with a periodic term.

On the Fig.~\ref{fig:PATCL} middle panels, the rate differences multiplied by the time elapsed since J2000  (from 1970 to 2050) give an upper bound on the accumulated angle difference of 2.73\,mas for $\psi$ and 1.05\,mas for $\theta$ respectively --- corresponding to equatorial surface displacements of about 2.30\,cm and 0.87\,cm. Because $\Delta\psi$ and $\Delta\theta$ oscillate about zero, the actual accumulated differences remain bounded, well below these values. The free rotation angle, by contrast, shows a clear linear secular trend, from about $-0.56$\,arcsec to $+1.0$\,arcsec over 1970--2050, corresponding to an additional drift of about 11.7\,mas/yr on top of the remaining periodic signature. The integrated differences (bottom panels) remain far below these bounds. $\Delta \psi$ and $\Delta\theta$ stay within 0.019 mas and 0.007 mas. $\Delta \phi$, by contrast, drifts by 11.8~mas/yr (9.9~cm/yr at the equator), from about -0.35~arcsec in 1970 to +0.59~arcsec in 2050, with a leading  (annual) periodic term of 0.91~mas (8~mm) (see inset panel on Fig~\ref{fig:PATCL}. Both values agree with the first-order estimates of Table~\ref{tab:trc-first-order} to better than 1\%, which validates the conversion of Eq.~(\ref{eq:PATCL}).

For Mars, \cite{Baland2023} estimate that the  time-rotation coupling affects also Mars's rotation angle and rate  at up to 167\,mas (2.7\,m at the surface) and 7.3\,mas/day with an error relative to \cite{Konopliv2020} of about 9\,mas (15\,cm) induced by incomplete relativistic modeling. Reproducing this estimate directly with \tempus, by converting a Mars orientation series from TDB to TCM using the same methodology applied to the Moon (Eq. \ref{eq:PATCL}), would provide an independent validation of this result; we plan to carry out this computation in future work.

For Moon and Mars, time-rotation coupling terms are  comparable to current tracking precision; if unmodeled, they could plausibly bias the parameters these data are used to fit.

For Mercury and Venus, rotation itself has been studied only within Newtonian dynamics~\citep{Noyelles2013,Etienne2025, Phan2025} or, in a relativistic spin-orbit model, unvalidated against spacecraft data~\citep{RambauxBois2004}, and not even the time-rotation coupling quantified for the Moon and Mars has been computed for these planets.

The amount of the time-rotation coupling in the rotation angle $W$ follows at first order from the rotation rate $\dot W$ and the offset $\Delta t=\mathrm{TCX}-\mathrm{TDB}$ between the body's own time scale and TDB: $\delta W \simeq \dot W\cdot\Delta t$. Since TDB falls behind TCB at the constant rate $L_B=1.55\times10^{-8}$ and TCX at the mean rate $L_X$ listed as the TCB$-$TCX drift in Table~\ref{tab:klioner}, $\Delta t$ changes at the rate $(1-L_X)/(1-L_B)-1 \simeq L_B-L_X$, neglecting terms of order $10^{-16}$; thus $\delta W$ approximately drifts at the rate $\dot W\cdot(L_B-L_X)$. The difference $L_B-L_X$ is positive for the Moon, Mars and the Galilean satellites, whose clocks run faster than TDB, and negative for Mercury and Venus. The dominant periodic part of $\delta W$ is the term at the orbital period of the body (or of its host planet): integrating Eq.~(\ref{eq:alpha}) along a Keplerian orbit around the Sun, of semi-major axis $a$ and eccentricity $e$, gives to first order in $e$ an oscillation of amplitude $2\sqrt{GM a}\,e/c^2$ in $\Delta t$, hence $\dot W\cdot 2\sqrt{GM a}\,e/c^2$ in $\delta W$. With these expressions we reproduce the 11.7\,mas/yr found above for the Moon to within 1\% and the values given by \cite{Baland2023} for Mars, Mercury and Venus, with periodic amplitudes agreeing to within 1\% as well. Table~\ref{tab:trc-first-order} gives the resulting values. They show that the size of the effect depends on the product of the rotation rate and the clock offset, not on the clock offset alone. Mercury's time scale departs from TDB four times faster than that of Mars, but Mercury rotates 57 times more slowly, so the orientation correction is smaller. Conversely, the Galilean satellites, which rotate once per orbit in only 1.8 to 16.7 days, reach the decimetre-to-metre level of displacement at the equator. For Mercury, the periodic term (3.2\,mas) is comparable to the geodetic nutation (5.08\,mas; \citealt{kgs2009}), whereas for Mars the time-rotation coupling dominates. The ordering of the relativistic rotation terms therefore changes from body to body.

These estimates show how large the effect is, not whether it can be detected. The drift term in particular is absorbed by the model's initial conditions fitted to the observations, so only the periodic term has to be compared with the rotation solutions, which are expected to improve significantly with BepiColombo for Mercury and with JUICE for the Galilean satellites, Ganymede in particular. For Mercury, simulations of the BepiColombo rotation experiment give a precision of 1.4\,arcsec on the annual libration amplitude under worst-case assumptions \citep{pfyffer2011}, and of 0.1--0.8\,arcsec, depending on the scenario, in full simulations with tracking, accelerometer and camera data \citep{Cicalo2016}. This is still 30 to 430 times above the 3.24\,mas of Mercury's periodic term, so its time coupling is currently neglectable. Applying Eq.~(\ref{eq:PATCL}) with the \tempus time scales to planetary bodies and moons would go beyond the first-order estimates of Table~\ref{tab:trc-first-order}; which we leave to future work.

\begin{table}[htbp]
\caption{First-order estimations for time-rotation coupling in the rotation angle (see text): secular rate and amplitude of the main periodic term. For the Galilean satellites, $L_X$ is approximated by Jupiter's drift in Table~\ref{tab:klioner} plus $1.5\,GM_J/(a\,c^2)$, for a circular orbit of radius $a$ around Jupiter. Rotation rates used are from \citet{Archinal18}. In parentheses, the displacement at the equator (per year for the secular term).}
\label{tab:trc-first-order}
\centering
\small
\begin{tabular}{lrrr}
\hline\hline
Body & $\dot W$ & Secular & Periodic \\
     & [$^\circ$\,d$^{-1}$] & [mas\,yr$^{-1}$] & [mas] \\
\hline
Moon     & 13.18   & 11.8 (9.9\,cm) & 0.91 (8\,mm) \\
Mars     & 350.89  & 2669.5 (43.95\,m) & 167.12 (2.75\,m) \\
Mercury  & 6.14    & 183.6 (2.17\,m) & 3.24 (4\,cm) \\
Venus    & $-1.48$ & 9.7 (28\,cm) & 0.04 (1\,mm) \\
Io       & 203.49  & 2046.1 (18.15\,m) & 92.79 (82\,cm) \\
Europa   & 101.37  & 1267.6 (9.60\,m) & 46.22 (35\,cm) \\
Ganymede & 50.32   & 706.9 (9.02\,m) & 22.94 (29\,cm) \\
Callisto & 21.57   & 327.2 (3.82\,m) & 9.84 (11\,cm) \\
\hline
\end{tabular}
\end{table}

\subsection{Time scales for lunar and Martian navigation}
\label{sec:app:missions}

The application that motivated the advent of \tempus is the Position, Navigation and Timing (PNT) service, and in particular, the Lunanet constellation at the Moon \citep{LNISv5} and its ESA Martian counterpart MARCONI \citep{Melman24}, in whose preparatory study \tempus was developed. A user derives one-way ranges from the constellation's clocks, which are kept on a system time traceable to Earth, so an error in the transformation between the body's coordinate time and TT enters the user's timing directly, at about $30$\,cm of range per 1\,ns.

For the Moon the IAU has defined TCL \citep{IAU24}; LunaNet requires every provider to deliver timing synchronised to a standard lunar time, or the corrections needed to reach it \citep{LNISv5}; and the architectures studied for ESA Lunanet contribution (Moonlight) tie the satellite clocks to the constellation reference at the $1$--$5$\,ns level, within a $10$\,m signal-in-space budget \citep{Iess25,Sesta25}. MARCONI aims to offer the same kind of one-way ranging at Mars, with expected accuracies of $15$\,m for a surface user aided by a digital elevation model, below $100$\,m for a lander and $80$\,m or better for an orbiter \citep{Melman24}. The transformation these systems have to undergo is not small. TCL runs slower than TCG by $1.48\,\mu$s per day (from Table~\ref{tab:klioner}), with periodic terms of up to $0.47\,\mu$s from the eccentricity of the lunar orbit \citep{Turyshev25}, so a nanosecond is $0.2\%$ of that amplitude; TCM runs faster than TCG by $0.44$\,ms per day.

Four sources contribute to the error of a computed series, each bounded by one of the comparisons of Sects.~\ref{sec:validation} and \ref{sec:perturbers}, and they span four orders of magnitude. 
Largest is the truncation of the model itself: the $\beta$ term of Eq.~(\ref{eq:dtcx_dtcb}) contributes $1.1\times10^{-16}$\,s\,s$^{-1}$ (block~(a) of Table~\ref{tab:ephemerides}), or $3.5$\,ns per year, so a transformation carried only to $c^{-2}$ with the $\alpha$ contribution only leaves the requirement within the year. 
Next is the completeness of the dynamical model: with all $373$ small bodies of DE440 supplied, \tempus still differs from LTE440 by $1.6\times10^{-17}$\,s\,s$^{-1}$, or $0.5$\,ns per year (Table~\ref{tab:validation}), which is the part of DE440's lunar model that \tempus does not carry. 
Smaller again is the impact of choosing different ephemerides, up to $1.9\times10^{-18}$\,s\,s$^{-1}$ for the Moon and $2.4\times10^{-18}$ for Mars, below $0.1$\,ns per year, though with detrended peaks reaching $1.7$\,ns over $200$\,yr (block~(c); Sect.~\ref{sec:perturbers:ephem}). 

The smallest possible error is from \tempus numerical integration itself, shown by the comparison with the TU~Dresden solution: both use the same INPOP19a ephemerides on the same body set, so only their implementations differ by at most $6.9$\,ps for TCL and $4.9$\,ps for TCM (Table~\ref{tab:klioner}) for the peak difference.

Therefore, against $1$--$5$\,ns, only the lunar dynamical model, at half a nanosecond per year, is relevant for how long a computed series stays inside the requirement. The distinction between the drift and the max peaks of the residual, properties given in all Tables, matters here: a drift is in part absorbed by the periodic re-synchronisation such a system can perform, whereas the residual peaks are not. A constellation time scale should in any case be computed from the same ephemerides that produce its satellite orbits. Then there is no mismatch between the two, and the completeness of the dynamical model is left as the term to improve.

%% file: sections/conclusions.tex
%
%

\section{Conclusions}
\label{sec:conclusions}

With \tempus we provide a validated, ephemeris-agnostic implementation of the IAU 2000 time transformation, Eq.~(\ref{eq:dtcx_dtcb}), for any solar-system body at its center of mass. TCX$-$TCB is integrated numerically from any available ephemerides  (INPOP, DE or EPM) in their TDB or TCB releases, with dynamical modeling scalable by the user, and is tabulated against the time scale of that ephemerides or, as the TU~Dresden time solutions do \citep{Klioner}, against TCX itself (Sect.~\ref{sec:coords:argument}), written directly or as a Chebyshev polynomial product. Every run is reproducible from a single command line.

The implementation of Eq.~(\ref{eq:dtcx_dtcb}) has been validated against the TU~Dresden time solutions \citep{Klioner} and the TCG$-$TCB solution distributed with INPOP19a: over a $200$\,yr span the drifts agree at the $10^{-21}$\,s\,s$^{-1}$ level and the detrended residuals stay below $25$\,ps (Tables~\ref{tab:klioner} and \ref{tab:validation}), far below the accuracy of any current atomic clock. The comparison with LTE440 is limited instead by the remaining difference in the lunar dynamical model, not by the integration of Eq.~(\ref{eq:dtcx_dtcb}) itself.

The five largest main-belt asteroids account for some $60\%$ of the $4.3$--$4.7\times10^{-18}$\,s\,s$^{-1}$ by which the full main belt shifts TCX$-$TCB for the Earth, Mars and the Moon; the trans-Neptunian objects add a further $2$--$3\times10^{-18}$; and the choice of ephemerides contributes up to $2.4\times10^{-18}$.

A body time scale is therefore only as good as its consistency with the ephemerides and body set that produced the trajectories it serves, and \tempus makes that consistency a matter of passing the correct mass and trajectory files. 

\tempus has already been used for the ILuRS2026 lunar products and in ESA's MARCONI study. Against the nanosecond synchronisation that lunar and Martian navigation constellations require, neither our implementation of the numerical integration nor the choice of input ephemerides is the limiting term; the completeness of the dynamical model is. It also supplies the time argument in which planetary rotation must be modelled (see Sec.~\ref{sec:app:rotation}).

The \tempus software will be made available as open source.
Natural future extensions would be the introduction of the perturbations induced by a ring, support and tests of natural satellite ephemerides with the parent planet's oblateness for bodies such as Ganymede and Titan, and a surface time output option referred to a chosen geoid value.